\documentclass[manuscript]{acmart}

\AtBeginDocument{%
  }

\begin{document}

\title[Polymarket]{Prediction Laundering: The Illusion of Neutrality, Transparency, and Governance in Polymarket}

\author{Yasaman Rohanifar}
\email{yasamanro@cs.toronto.edu}
\affiliation{%
  \institution{Computer Science, University of Toronto}
  \state{Ontario}
  \country{Canada}
}

\author{Syed Ishtiaque Ahmed}
\email{ishtiaque@cs.toronto.edu}
\affiliation{%
  \institution{Computer Science, University of Toronto}
  \state{Ontario}
  \country{Canada}
}

\author{Sharifa Sultana}
\email{sharifas@illinois.edu}
\affiliation{%
  \institution{University of Illinois Urbana-Champaign}
  \state{Illinois}
  \country{USA}
}

\begin{abstract}
The growing reliance on prediction markets as epistemic infrastructures has positioned platforms like Polymarket as providers of objective, real-time probabilistic truth. However, the authoritative signals produced by these platforms often obscure the underlying uncertainty, strategic manipulation, and capital asymmetries inherent in their production, leading to an inappropriate epistemic trust by external observers. This paper presents a qualitative sociotechnical audit of Polymarket ($N=27$), combining digital ethnography with interpretive walkthroughs and semi-structured interviews to examine how probabilistic authority is enacted and contested. We introduce the concept of "Prediction Laundering", drawing on MacFarlane’s framework of knowledge transmission, to describe the sociotechnical process by which subjective, high-uncertainty bets, strategic hedges, and capital-heavy whale movements are scrubbed of their original noise and personal stakes through algorithmic aggregation. We trace this process through a four-stage laundering lifecycle that begins with "Structural Sanitization", where a centralized ontology scripts the boundaries of the bet-able future. This is followed by "Probabilistic Flattening", where the interface collapses heterogeneous motives. The third stage, "Architectural Masking", utilizes selective opacity to hide capital-driven influence behind a mirage of crowdsourced consensus. Finally, the process concludes with "Epistemic Hardening", where messy governance disputes are erased to produce an objective historical fact. Our findings reveal that this process induces a state of epistemic vertigo among users and creates significant accountability gaps by offloading the labor of truth-resolution to off-platform communities like Discord. We challenge the narrative of frictionless collective intelligence and demonstrate that the platform produces a state of Epistemic Stratification, where technical elites audit the underlying data while the general public consumes a sanitized, capital-driven signal. We conclude by arguing for a shift toward Friction-Positive Design, advocating for interfaces that expose rather than erase the social and financial friction inherent in the production of synthetic truth.
\end{abstract}


\keywords{Prediction Markets, Digital Ethnography, Information Markets}

\maketitle
\section{Introduction}

Prediction markets have gained renewed global attention as computational systems capable of aggregating dispersed beliefs into probabilistic forecasts \cite{massa2025kalshi, cross2026prediction}. From high-stakes geopolitical conflicts to public health and economic instability, market prices are increasingly cited in media, policy discussions, and online publics as reliable indicators of future events \cite{andersen2024bloomberg, Miah2025Media, tapscott2024biden}. Platforms such as Polymarket \cite{polymarket}, operating via blockchain infrastructure and pseudonymous participation, are framed as uniquely credible due to their perceived transparency and decentralized financial incentives \cite{yahoofinance2025polymarket}.

However, this growing reliance raises a fundamental question for the fairness, accountability, and transparency (FAccT) community: what kind of epistemic authority do these markets actually exercise, and what risks are hidden behind their probabilistic outputs? While prior research has focused on forecasting accuracy and market efficiency, far less attention has been paid to these systems as sociotechnical infrastructures that actively shape knowledge, belief, and responsibility. Treating probabilistic prices as neutral or objective obscures the ways in which platform design, governance rules, and social dynamics influence whose beliefs count, which futures are legible, and how uncertainty is morally framed.

In this paper, we argue that prediction markets function through a process we term \textbf{"Prediction Laundering"}. Drawing on MacFarlane’s work on knowledge transmission \cite{macfarlane2005making,macfarlane2014assessment}, we define Prediction Laundering as the sociotechnical process by which individual, high-uncertainty bets and strategic hedges are scrubbed of their original noise and personal stakes through market aggregation. This process flattens the "information fog" and opaque governance experienced by participants into a high-fidelity percentage, recirculating it as objective, de-contextualized knowledge.

To investigate these dynamics, we conduct a qualitative sociotechnical audit of Polymarket, combining critical digital ethnography with interpretive walkthroughs and semi-structured interviews ($N=27$). We move beyond accuracy metrics to examine the visible and invisible walls of the system, investigating how probabilistic authority is enacted through design, market rules, community practices, and off-platform coordination. Our research is guided by three core questions:

\begin{itemize}
    \item \textbf{RQ1:} How do prediction market design and pricing mechanisms transform the users' contested beliefs and value-laden judgments into seemingly neutral probabilistic signals?
    \item \textbf{RQ2:} How do users interpret, rely on, and contest these probabilistic signals, particularly when moral discomfort, uncertainty, or exclusion arise?
    \item \textbf{RQ3:} What kind of risks and challenges occur for users when reality, rumors, and probabilities as numbers convolute together?
\end{itemize}

We show that Polymarket curates a narrow ontology of the future, privileging capital-driven influence over diverse belief aggregation while offloading intensive interpretive labor onto participants. Our findings trace this through a four-stage laundering lifecycle that reveals systemic failures in epistemic governance:

\begin{enumerate} 
    \item \textbf{Structural Sanitization:} The platform utilizes a centralized ontology to script which versions of the future are bet-able, pre-formatting reality for speculation. 
    \item \textbf{Probabilistic Flattening:} The interface collapses heterogeneous motives, such as strategic hedging and interpretive oscillation, into a singular numerical truth. 
    \item \textbf{Architectural Masking:} Selective opacity hides capital concentration and whale influence, creating a mirage of democratic consensus.
    \item \textbf{Epistemic Hardening:} The dirty labor of governance disputes and moral discomfort is erased to produce de-contextualized, authoritative facts.
\end{enumerate}

By identifying the \textbf{Epistemic Stratification} and \textbf{Accountability Gaps} created by this process, this work challenges optimistic narratives of collective intelligence. We call for greater scrutiny of the computational systems mediating how synthetic truths are imagined, priced, and trusted.
\section{Related Work}

To understand the sociotechnical mechanisms of prediction markets, we must engage with literature that spans epistemology, economic sociology, and algorithmic accountability. This section establishes the theoretical foundations for our audit by examining three core areas. First, we explore the philosophy of "Epistemic Transmission" and the concept of "laundering" to understand how information is transformed as it moves between contexts. Second, we analyze the "Wisdom of Crowds" as an idealized framework for collective intelligence and contrast it with the realities of market stratification. Finally, we situate this study within the field of algorithmic accountability, drawing on sociotechnical auditing methods to investigate the governance and labor required to maintain these informational systems.

\subsection{Knowledge Laundering and Epistemic Transmission}
The primary theoretical anchor for this paper is the concept of "Knowledge Laundering". This term was coined by John MacFarlane to describe a fundamental failure in the transmission of information between distinct contexts. In the framework established by MacFarlane \cite{macfarlane2014assessment, macfarlane2005making}, knowledge is not a fixed or invariant property. Instead, it is often assessment sensitive, meaning its validity depends on the epistemic stakes and the specific context of the person evaluating a claim. The process of laundering occurs when a piece of information moves from a high stakes context where a subject might be cautious or uncertain to a low stakes context where an external observer accepts the claim as a settled fact. This transition happens without the underlying uncertainty being transmitted along with the information itself. 

This theory challenges the "Transmission Principle" in classical epistemology which traditionally assumes that if Person A knows $p$ and tells Person B, then Person B also knows $p$ \cite{lackey2008learning}. MacFarlane argues that this principle breaks down when the pedigree of the knowledge is scrubbed. If Person A only knows $p$ relative to a very low standard of evidence but Person B accepts $p$ as an objective and high standard fact, the knowledge has been laundered. This creates a loophole where the public can end up trusting a claim that no single individual in the chain actually felt certain about.

In the context of Algorithmic Accountability, this framework provides a powerful lens for examining how computational systems mediate belief. While MacFarlane originally applied this to human testimony, we argue that prediction markets act as automated laundry machines for epistemic claims. Current FAccT scholarship on inappropriate reliance \cite{metaxa2021auditing} and algorithmic opacity \cite{burrell2016machine} has largely focused on how users over-trust model outputs. However, by integrating MacFarlane’s work, we shift the focus to the transformation of the signal itself. We contend that prediction markets do not just display information. They perform a sociotechnical scrubbing that converts strategic and capital driven bets into de-contextualized probabilistic truth. This process enables what we define as "Prediction Laundering", referring to the systemic removal of the dirty context involving hedges, noise, and power asymmetries from the final public signal.

\subsection{Information Markets and the ``Wisdom of Crowds''}
The conceptual foundation for prediction markets rests on the Hayekian view of markets as a unique solution to the problem of dispersed knowledge \cite{hayek2013use}. Hayek argued that the price mechanism allows society to aggregate fragmented bits of information that no single central planner could possess. In this view, prices function as signals that coordinate human action by summarizing vast amounts of local and specialized data. Modern proponents of prediction markets extend this logic by suggesting that these platforms are the most efficient tools for "discovery" in uncertain environments. Wolfers and Zitzewitz formalize this in their analysis of prediction markets in theory and practice, stating that the financial incentives of the market force participants to reveal their true beliefs \cite{wolfers2004prediction}. They argue that this makes market prices more reliable than polls or expert opinions because the "skin in the game" requirement filters out insincere or low-quality information.

Standard economic interpretations of prediction markets are primarily rooted in the "Wisdom of Crowds" hypothesis. This framework suggests that a large group of independent actors will collectively produce a more accurate forecast than any individual expert \cite{surowiecki2005wisdom}. The core assumption is that a market mechanism can effectively filter out individual biases and noise by providing financial incentives for accuracy. Wolfers and Zitzewitz argue that this process makes prediction markets uniquely efficient at information aggregation because participants are forced to back their beliefs with capital \cite{wolfers2004prediction}. In this idealized model, the market price is viewed as a neutral and mathematical consensus that reflects the total sum of available public and private information.

However, these Hayekian and economic ideals assume a level of democratic participation that is rarely present in contemporary technical infrastructures. Sociological critiques of finance suggest that markets are better understood as specialized micro-structures rather than broad and egalitarian crowds. For example, Knorr Cetina notes that global financial markets are deeply stratified environments where information is mediated by social networks and technical expertise \cite{knorr2002global}. In these settings, a small subset of participants with high capital and specialized platform knowledge exerts a disproportionate influence over the resulting signal, creating a state of specialized competition where the price does not reflect a collective belief but rather the strategic moves of a few dominant actors. We build on these critiques by examining how the "crowd" in decentralized platforms like Polymarket is structured. We focus on the gap between the platform's presentation as an open and egalitarian marketplace and the reality of its capital-driven hierarchy.
\vspace{-0.5cm}
\subsection{Algorithmic Accountability and Sociotechnical Audits}
The study of algorithmic accountability has shifted from a narrow focus on mathematical transparency toward an investigation of the sociotechnical "assemblages" that sustain digital systems \cite{ananny2018seeing}. Within this framework, platforms are not merely neutral hosts for data, but active governors that curate which futures are "speakable and wager-able". Our methodology follows the tradition of "auditing from the outside", which uses qualitative user experiences to reverse-engineer the social consequences of opaque systems \cite{sandvig2014auditing}. This approach acknowledges that knowledge is inherently situated in people and locations, and that the "wisdom" produced by a platform is inseparable from the social dynamics of the community that generates it. By focusing on the lived experiences of participants, we can observe the emotional and cognitive negotiations that are typically rendered invisible by raw transaction data and market charts.

A critical dimension of this accountability is the "invisible labor" required to maintain the appearance of a functioning, objective market. Drawing on Jackson’s "Broken World Thinking", we argue that the credibility of prediction markets often depends on constant repair work performed in off-platform spaces like Discord, Telegram, or Twitter \cite{jackson2014rethinking}. While these platforms are marketed as decentralized and automated, their governance is frequently characterized by resource dependencies and asymmetrical power relations between platform owners and users. This resonates with broader anthropological critiques of digital technology, which suggest that infrastructures often reinforce existing inequalities while creating new forms of exclusion under the guise of democratization. In the case of prediction markets, the platform captures the value of a "clean" probabilistic signal while offloading the "dirty" work of dispute resolution and fact-checking onto a pseudonymous public that lacks formal participation rights.

Finally, we address the risk of "inappropriate reliance", where external observers or policymakers treat market outputs as de-contextualized facts \cite{metaxa2021auditing}. This reliance is encouraged by a platform design that prioritizes "stickiness", cognitive ease, and social proof, often at the expense of epistemic rigor. Research on financial psychology shows that factors like loss aversion and framing effects create systematic patterns in how users navigate these environments, yet these insights are rarely applied to the governance of forecasting infrastructures. By identifying the mechanisms of Prediction Laundering, we bridge the gap between behavioral economics and platform studies. We argue that the lack of accountability in these systems is not a technical bug but a structural feature that allows speculative infrastructures to operate independently of traditional status markers or ethical constraints.
\section{Background}

Markets have long been understood in sociotechnical research as complex assemblages embedded in cultural, social, and material practices rather than mere arenas of economic exchange. Traditional and informal markets unfold through situated practices of negotiation, trust, and testing that constitute them as lived social worlds rather than mere price-setting mechanisms. In recent years, digital technologies have given rise to decentralized prediction markets like Polymarket, which transform future uncertainty into financial instruments. Polymarket achieves this information aggregation through the trading of binary event contracts, or "outcome tokens", which pay \$1 if a specified event occurs and \$0 otherwise. The trading price, fluctuating between \$0 and \$1, is treated as a collective assessment of probability; for instance, a contract trading at \$0.75 signals a 75\% market-estimated likelihood of the event.

The platform's infrastructure integrates three critical technological components that facilitate this epistemic function. First, Automated Market Makers (AMMs) replace traditional order books with mathematical formulas that supply pricing and liquidity for continuous trading. Second, the system utilizes blockchain settlement on the Polygon network, where all trades and transactions are recorded immutably on-chain using the USDC stablecoin. Finally, event resolution is handled by decentralized oracles that deliver verified real-world data to smart contracts, enabling automatic and tamper-resistant settlement. This technical stack ensures that the process of "finding the truth" is tied to a rigid, automated logic that bypasses traditional institutional intermediaries.

A defining characteristic of this infrastructure is pseudonymous participation via blockchain wallet addresses. This design fosters a paradoxical environment where radical transparency of trading history coexists with strong privacy protections. While these mechanisms promise efficient aggregation, the platform effectively curates an ontology of the future by permitting only uncertainties that are legible, measurable, and deemed morally acceptable by the platform's governance. By privileging speed over accuracy through its automated architecture, the system does not merely reflect the future but actively structures how it can be financially and epistemically engaged.
\section{Methods}

We conducted a qualitative sociotechnical audit of Polymarket to examine how probabilistic authority is enacted, interpreted, and contested through market rules and community practices. Polymarket was selected as the primary site of study because it is the leading global, blockchain-based prediction platform. Given the platform's technical complexity and pseudonymous culture, we utilized participant interpretation and facilitated engagement to capture the lived reality of the system without necessitating real-world monetary risk.

\begin{table}[!t]
\centering
\begin{tabular}{ |c|c|c|c|c|c|c|c| }
    \cline{1-8}
    \multicolumn{8}{|c|}{\textbf{Total: 27 (Female: 11 , Male: 16)}} \\
    \cline{1-8}
    \multicolumn{2}{|c|}{\textbf{Age Range}} &  
    \multicolumn{2}{c|}{\textbf{Education}} &  
    \multicolumn{2}{c|}{\textbf{Crypto Familiarity}} &
    \multicolumn{2}{|c|}{\textbf{Expertise}} \\
    \cline{1-8}
    
    \multicolumn{1}{|r}{20-29:} & \multicolumn{1}{l|}{8}  &  
    \multicolumn{1}{|r}{Highschool:} & \multicolumn{1}{l|}{1} & 
    \multicolumn{1}{|r}{Novice:} & \multicolumn{1}{l|}{10} &
    \multicolumn{1}{r}{Domain Expert:} & \multicolumn{1}{l|}{10} \\
    
    \multicolumn{1}{|r}{30-39:} & \multicolumn{1}{l|}{9}  &  
    \multicolumn{1}{|r}{Undergraduate:} & \multicolumn{1}{l|}{12} & 
    \multicolumn{1}{|r}{Exchange-Reliant:} & \multicolumn{1}{l|}{11} &
    \multicolumn{1}{r}{Generalist User:} & \multicolumn{1}{l|}{17} \\
    
    \multicolumn{1}{|r}{40-49:} & \multicolumn{1}{l|}{5}  & 
    \multicolumn{1}{|r}{Masters:} & \multicolumn{1}{l|}{9} &
    \multicolumn{1}{|r}{Crypto-Native:} & \multicolumn{1}{l|}{6} &
    \multicolumn{1}{r}{} & \multicolumn{1}{l|}{} \\
    
    \multicolumn{1}{|r}{50+: } & \multicolumn{1}{l|}{5}  & 
    \multicolumn{1}{|r}{Doctorate:} & \multicolumn{1}{l|}{5} & 
    \multicolumn{1}{|r}{} & \multicolumn{1}{l|}{} &
    \multicolumn{1}{r}{} & \multicolumn{1}{l|}{} \\
    
    \cline{1-8}
\end{tabular}
\captionsetup{justification=centering}
\caption{Demographics of Participants}
\label{participants-info}
\vspace{-6mm}
\end{table}

\subsection{Multi-Modal Qualitative Approach}
We employed a multi-modal qualitative approach over a four-month period to capture the platform as a sociotechnical assemblage. This design moved from macro-level platform observations to micro-level individual negotiations of truth and power. The study began with Phase 1, which focused on a Digital Ethnography where we conducted systematic observation of platform dynamics across the primary interface and its peripheral social ecosystems. This phase consisted of three sub-components including a technical audit of Polymarket affordances, the monitoring of social shadow infrastructures on X, Reddit, and Discord, and the tracking of high-capital whale movements via block explorers. In this stage, we adopted a non-participant observer role to archive field notes, screenshots, and transaction data without intervening in the social dynamics.

Following the ethnography, we conducted one-on-one sessions with the full sample of participants ($N=27$) in Phase 2. Each session integrated three distinct methods to ensure a holistic understanding of the user experience. The engagement began with Interpretive Walkthroughs using think-aloud protocols while navigating real market interfaces. This stage focused on immediate and visceral reactions while capturing the real-time reasoning process as users encountered specific event contracts and price fluctuations.

The session then moved into Semi-Structured Interviews investigating epistemic entry points. Rather than focusing solely on historical trust, these interviews explored the specific technical, social, and aesthetic cues participants use to evaluate the legitimacy of decentralized information. We documented the criteria users employ to transition from passive observers to active interpreters of speculative data, focusing on how they bridge existing knowledge with the probabilistic outputs generated by the system.

The final part of the session utilized Design Probes where participants were confronted with specific artifacts captured during the initial digital ethnography. These probes included archived comment threads, screenshots of significant whale positions that influenced market pricing, and ethical vignettes involving sudden market volatility. By presenting these real-world fragments of platform life, we prompted participants to vocalize their internal negotiations and tactical responses to the social and technical friction found within the ecosystem. This integrated pipeline ensured that the environmental data from Phase 1 directly informed the individual-level inquiries in Phase 2, allowing us to observe how users navigate the gap between the clean signal of the interface and its messy social reality.

\begin{figure*}[!t] 
    \centering 
    \includegraphics[width=0.7\textwidth]{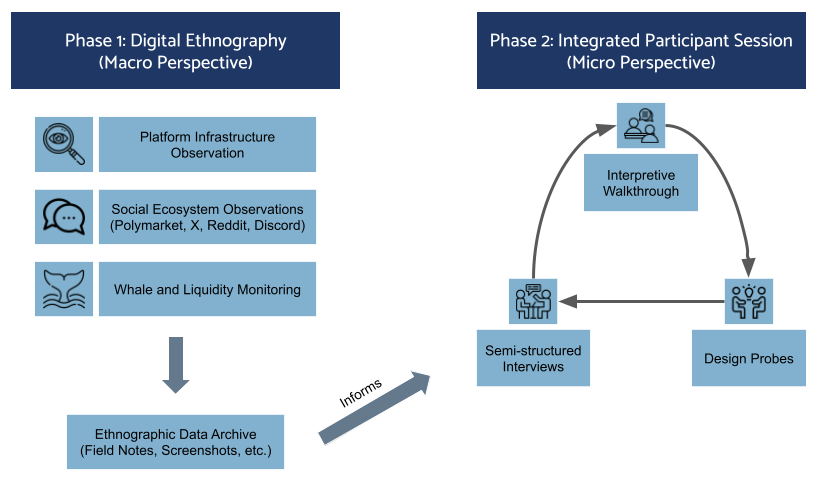} 
    \caption{Methodological workflow and data integration diagram illustrating our two-phase research design moving from macro-level platform ethnography to micro-level participant sessions. The Ethnographic Data Archive from Phase 1 directly informs the Design Probes in Phase 2 to enable a confrontation between users and real-world platform artifacts.}
    \captionsetup{skip=2pt}
    \label{methods_workflow} 
\end{figure*}
 
\subsection{Participants and Sampling}
We employed a hybrid purposive and snowball sampling strategy to recruit twenty-seven participants ($N=27$) representing a diverse range of backgrounds and technical proficiencies. To ensure the audit captured a wide spectrum of information literacies, we seeded our recruitment through professional and academic networks, targeting individuals with varying degrees of exposure to financial forecasting and decentralized technologies. We categorized these participants into two primary groups based on their professional and academic backgrounds. "Domain Experts" ($n=10$) consist of individuals with advanced training in computer science, economics, or data analysis. These participants possessed the theoretical and technical background necessary to interrogate the underlying blockchain protocols and the mathematical logic of the platform's Automated Market Makers. Conversely, "Generalist Users" ($n=17$) represent individuals from diverse professional fields such as the arts, education, and healthcare who utilize digital platforms but lack specialized training in predictive modeling or decentralized finance.

Beyond formal expertise, we further stratified our sample by crypto familiarity, as the ability to navigate the platform’s blockchain infrastructure significantly impacts a user’s power position within the ecosystem. We define "Crypto-Native" participants as those who regularly interact with self-custodial wallets and decentralized protocols, possessing the literacy to audit transactions via block explorers. "Exchange Reliant" users are those familiar with centralized cryptocurrency exchanges who understand the value of digital assets but lack experience with on-chain governance or smart contract interaction. Finally, "Novice" participants include individuals with little to no prior exposure to cryptocurrency, for whom the platform’s blockchain-based settlement remains an opaque black box. By cross-referencing these categories, we established a participant matrix that allowed us to observe how power asymmetries manifest differently for those who can see the underlying mechanisms versus those who must rely on the platform’s front-end presentation.

Following these initial sessions, we utilized snowball sampling \cite{goodman1961snowball, noy2008snowball} to reach participants who occupy specific outsider niches, such as individuals with strong moral or religious objections to speculative markets. This approach allowed us to move beyond a convenience sample and intentionally include voices that are systematically excluded or marginalized by the platform's technical jargon and complex crypto infrastructure. By including these epistemic outsiders, we were able to document the barriers to entry and the moral friction that characterize the experience of those for whom the market’s logic is not inherently intuitive. This inclusive sampling strategy ensures that our audit accounts for the lived experiences of marginalization and exclusion that often remain invisible in purely quantitative analyses of market efficiency.

\subsection{Analysis Procedure}
All sessions were recorded, transcribed, and analyzed through iterative thematic coding, moving from initial open codes to more structured conceptual categories. Our analytical framework was primarily abductive, meaning we treated moments of breakdown, irony, and contestation as critical entry points into the ways participants and traders make sense of uncertainty. Rather than seeking a unified user experience, we intentionally focused on the friction between the platform's technical affordances and the users' situated knowledge.

This approach allowed us to identify the specific patterns of sociotechnical mediation that facilitate the translation of subjective belief into probabilistic signals. By centering power relations, we interrogated how the platform’s interface and governance structures privilege certain forms of expertise while marginalizing others. This method of "critical listening" ensured that our findings reflect not just the stated functionality of the platform, but the lived experiences of those navigating its speculative infrastructure. Ultimately, our analysis sought to connect individual moments of confusion or skepticism to broader systemic accountability gaps, providing a grounded basis for our later theorization of the platform's epistemic outcomes.
\section{Findings: The Life Cycle of a Laundered Prediction}

Our sociotechnical audit reveals that Polymarket’s primary function is the systematic "laundering" of subjective, value-laden, and often adversarial human inputs into objective-looking probabilistic signals. While the platform’s interface presents a veneer of mathematical transparency, this clarity is achieved through a series of structural and social filters that strip away the "dirty" context of individual wagers, including financial hedging, coordinated manipulation, and capital-driven bias. We trace this process through a "Laundering Lifecycle" and demonstrate how the platform’s mechanics consistently privilege capital and technical literacy over democratic information aggregation. By examining the transition from market creation to final settlement, we identify the specific points where the system collapses complex social disagreement into a sanitized numerical truth, effectively insulating the resulting data from accountability or ethical interrogation.

\subsection{Stage 1: Structural Sanitization (The Curated Ontology)}
The first stage of prediction laundering occurs before a single wager is placed, at the point where a real-world event is translated into a tradable contract. Our audit reveals that Polymarket does not function as a decentralized marketplace for all possible futures, but instead enforces a rigid, centralized ontology. While the platform encourages "market suggestions" via Discord and X, users cannot directly create or list their own markets \cite{polymarket_markets_2026}. Instead, a dedicated "Markets Team" acts as a primary epistemic filter, vetting proposals based on their "news value", trading demand, and technical resolve-ability. This structural gatekeeping ensures that only certain types of uncertainty are permitted to be quantified, effectively sanitizing the future before it ever reaches the user interface. One expert participant noted that this suggestion-based model creates a veneer of community participation while maintaining strict institutional control:

\begin{quote}
    ``I think people forget that Polymarket’s design actually shapes what you're even allowed to bet on. On paper, "anyone" can suggest creating a Polymarket! But in reality, the moderators have to approve the titles and how it's all going to be settled. If it’s too vague, like `Will the new iPhone be a success?' they'll reject it. Same for subjects that are too weird or [morally] bad... like markets on someone getting hurt. The result is this curated set of Polymarkets that they allowed for us... basically just the stuff that are measurable, legible, and fit their brand!'' (P3, Male, Crypto-native, Domain Expert)
\end{quote}

By filtering out the ambiguous or the socially unacceptable, the platform creates an epistemic boundary that privileges data-centric outcomes over complex, non-binary social realities. This curation suggests that the "wisdom" produced by the market is not a reflection of universal human speculation, but a subset of possibilities that have been pre-formatted for blockchain settlement. In the forthcoming discussion, we will further argue that the laundering process begins with this initial curation, where the platform decides which versions of reality are "fit" to be quantified.

\subsection{Stage 2: Probabilistic Flattening (The Collapsing of Belief into Price)}
Once a market is live, the platform utilizes a pricing mechanism that acts as an "epistemic solvent" by dissolving the specific context and motivations behind individual wagers. We observed that while participants entered the platform with a wide array of motivations, ranging from data-driven analysis to emotional bias, the interface consistently flattened these varied inputs into a singular numerical percentage. This mechanism collapses disagreement and moral contestation into a single signal that circulates as objective knowledge, effectively masking the messy reality of how that number was reached. One participant described the psychological shift that occurs when subjective opinions are converted into this clean, mathematical output:

\begin{quote}
    ``It's funny because once I see a number attached to a possible outcome, it stops feeling like an opinion and more like... real truth. It doesn't matter who is betting or why. You just see 72 percent and it feels like that is the truth, even though I know it is coming from... just a mess of guesses and incentives!'' (P9, Female, Exchange-Reliant, Generalist User)
\end{quote}

This flattening process often leads to a state of \textbf{"interpretive oscillation"} in which users fluctuate between treating the market as a credible tool and dismissing it as a gamified environment. Even as the interface presents high-fidelity data, participants reported a sense of epistemic vertigo, questioning if the price reflects information or merely platform-induced hype. One user expressed this confusion when reconciling the probabilistic framing with the underlying market mechanics:

\begin{quote} 
    ``I'm looking at all these percentages and I know it's supposed to be the probability... but there is this other language. The percentages don't add up. How much fee is taken by the platform? Is this number even realistic or just betting hype? What triggers the number to change? Like... does someone input it? I feel like I'm at a casino and everything is staged to make me bet!'' (P12, Male, Novice, Generalist User)
\end{quote}

The solvent effect is further complicated by \textbf{"the hedge problem"}, where the aggregator treats every dollar as conviction regardless of intent. Two participants described using the market for financial insurance, yet these wagers are indistinguishable from belief-driven bets on the front-end. This successfully launders financial hedging into an appearance of epistemic confidence, misleading the observer into believing the price reflects a consensus of belief rather than a cluster of conflicting financial strategies. Another participant explained the mechanics of this interpretive distortion when betting against their own preferences:

\begin{quote}
    ``If I were to bet on this election outcome, I'd bet against my own candidate just to soften the blow... if they lose, at least I get a payout! The problem is that other users see that I’m predicting a loss, when I’m actually just buying peace of mind... I wish there was a way to show which bets come from real belief and which are just financial hedges. These markets don't show if a politician is actually popular, but the media uses these numbers as the absolute truth!'' (P4, Male, Exchange-Reliant, Domain Expert)
\end{quote}

The transition from a mess of guesses to a singular percentage represents a fundamental loss of information. The platform strips away the "why" of the bet, leaving only the "what", which effectively misleads the observer into believing the price reflects a consensus of belief rather than a cluster of conflicting financial strategies. In our discussion, we explore this through the lens of \textbf{"Epistemic Dilution"}, where the market's need for a singular signal inevitably produces a distorted representation of public sentiment. This solvent effect ensures that by the time a user views the Polymarket landing page, the traces of hedging and doubt have been scrubbed away, leaving only the authoritative glow of the percentage.

\subsection{Stage 3: Architectural Masking (The Illusion of Crowdsourced Truth)}
Following the consolidation of price, the third stage of laundering involves obscuring the specific actors responsible for market movements. Although prediction markets are frequently framed as "democratic aggregators", our audit found that epistemic influence is disproportionately asserted through capital. The platform’s pseudonymity hides the identity of whales or large-scale traders whose massive positions can single-handedly shift the market—allowing their private financial interests to be laundered into the appearance of a broad, neutral consensus. This structural invisibility ensures that the final probabilistic signal appears to be a product of many minds rather than a few significant wallets. One participant noted how the weight of capital often overrides the wisdom of the general public:

\begin{quote}
    ``Obviously some people move the market way more than others. Just like real life, if you have money... or if you really care about a topic to lobby on it, or if you have insider information... your belief counts for more. Everyone else is kind of just reacting to that, not really contributing. The whale moves the price and then the crowd just follows along with that momentum... even if they don't know why it happened.'' (P17, Female, Exchange-Reliant, Generalist User)
\end{quote}

This asymmetry is maintained through a condition of \textbf{selective opacity}. While the primary Polymarket interface is clean and simple, it hides the complex on-chain data that reveals who is actually moving the money. We observed a significant literacy gap between crypto-native participants, who utilized third-party block explorers to audit whale behavior, and novice users, who were forced to take the front-end price at face value. This gap creates a two-tiered epistemic environment where the reality of market manipulation is legible only to technical insiders. A crypto-native participant described the necessity of looking under the hood to find the truth behind the price:

\begin{quote}
    ``To gain the most, you have to follow the whales. I'll just go to Polymarket  leaderboards, Etherscan or Arkham... [third-party tools] to see who is actually dumping millions into a vote. If you just look at the Polymarket homepage, it looks like many people are agreeing... but it's only when you dig deeper that you realize it is just one guy with a massive wallet!'' (P21, Male, Crypto-Native, Domain Expert)
\end{quote}

This selective opacity allows capital-driven signals to be laundered into the appearance of democratic agreement. By hiding the architects of the price movement, the platform prevents the average observer from questioning the motives behind a shift in odds. In our discussion, we connect this to the theory of \textbf{Epistemic Stratification}, where the ability to see through the laundry is a privilege reserved for those with specific technical and financial literacy, while the general public is left to consume the sanitized, consolidated output as fact.

\subsection{Stage 4: Epistemic Hardening (The Oracle Fallacy)}
The final stage of prediction laundering occurs during market resolution, where the "hardening" of a result erases the history of social, moral, and financial friction. This stage serves as a final sanitization by reclassifying ethically sensitive topics as neutral informational problems. By re-framing human tragedy such as war or political instability, as a probability number on a chart, the platform allows participants to shift the burden of moral responsibility onto the market itself. This objectification of harm normalizes speculation while insulating both the platform and the trader from ethical accountability. One participant described the discomfort of profiting from disaster and how the platform's design mitigates that guilt:

\begin{quote}
    ``There is a weird discomfort with betting on people dying in a war... but when you look at the chart, it just looks like any other stock market chart! The way all of the information is framed makes it feel like you are solving a puzzle rather than profiting from a disaster!'' (P2, Female, Exchange Reliant - Generalist User)
\end{quote}

By transforming a moral crisis into a speculative puzzle, the platform performs a form of moral laundering by taking the weight of a sensitive ethical choice and placing it on the mechanical neutrality of the market. This hardening ensures that once the oracle clears, the messy human cost of the wager is completely obscured. As another participant noted, the platform effectively absorbs the user's moral burden: ``it's like the platform will take the responsibility of being a bad person from you and place it on itself. You just do the betting!''

This moral distancing is paired with a tactical struggle to control the platform's resolution mechanisms. While the platform markets itself as a decentralized system, the actual process of resolving the truth relies on a stratified financial hierarchy. When a market outcome is ambiguous, the messy human labor of dispute resolution takes place in off-platform social silos like Discord, often involving tactical deception and succumbing to exit strategies. Once the market is settled by the oracle, however, this friction is scrubbed from the user interface, presenting the final result as an objective fact. One domain expert described the fundamental dishonesty that often precedes the hardening of a market signal:

\begin{quote}
    ``Polymarket is just a zero-sum game... and because of that, the comment section is probably one of the most intellectually dishonest places on the internet!! When you see someone being helpful or sharing a link, they almost always have a financial incentive to make you believe them. They want you to take the other side of their trade so they can get exit liquidity... they'll spam the record with news clips just to influence how the oracle eventually settles the truth.'' (P1, Female, Crypto-Native, Domain Expert)
\end{quote}

This tactical labor is eventually buried beneath the platform's formal resolution mechanics, which are shielded from the average user by high financial barriers. A participant with a technical background expressed concern over how this hardening process is governed:

\begin{quote} 
    ``What's weird is how these markets actually get resolved. On paper, it's democratic... but you have to stake a \$750 bond just to propose an outcome. Most people don't have that... you need capital just to have a say in how the truth is settled. Then, if you want to challenge, it's another \$750. People with the UMA tokens vote on it... so if you are wealthy, you can basically buy enough tokens to outvote everyone and make sure the market settles in your favor!'' (P16, Female, Exchange-Reliant, Domain Expert)
\end{quote}

The final "cleansing" of the record occurs at the moment of settlement. Another participant noted how the interface removes all traces of this chaotic behind-the-scenes labor once a result is reached:

\begin{quote}
    ``If a resolution is confusing... everyone runs to Discord to argue with the mods. It's total chaos behind the scenes. But then the Oracle clears... and all that mess is gone. On the main page, all that is left is a 'Yes' or 'No'... like there was never any doubt at all.'' (P25, Male, Crypto-Native, Domain Expert)
\end{quote}

The Oracle Fallacy lies in the public perception of the resolution as a neutral, data-driven event, when it is actually a contest of capital and strategic noise. In our discussion, we connect this to the \textbf{Accountability Gap}, where the platform takes credit for a "clean" output while offloading the "dirty" labor of dispute and the moral weight of the outcome onto the interface and third-party oracles.
\section{Discussion}

Our findings demonstrate that Polymarket’s epistemic authority is not a byproduct of neutral information aggregation, but the result of a deliberate sociotechnical scrubbing process we have termed \textbf{"Prediction Laundering"}. By tracing the lifecycle of a wager, we have shown how the platform systematically strips away the "dirty" context of individual motives such as financial hedging, capital-driven manipulation, and moral friction, to produce a sanitized, authoritative signal. This transformation creates a profound tension between the platform’s front-end veneer of "crowdsourced truth" and the back-end reality of capital concentration and technical gatekeeping. In the following sections, we discuss how this process induces a state of Epistemic Stratification and creates significant accountability gaps, ultimately challenging the idealized "Wisdom of Crowds" narrative that currently dominates the discourse on prediction markets.

\subsection{The Asymmetric Transparency of Synthetic Truth}
One of the cores promise of prediction markets is to solve the problem of "expert failure" by democratizing information aggregation. However, our findings suggest that Polymarket’s architecture does not distribute epistemic power but merely shifts it from traditional institutional experts to a new, invisible class of capital-rich technical elites. This results in what we call "Epistemic Stratification", where the platform produces a synthetic truth that appears democratic on the front-end but is technically and financially gated on the back-end.

This stratification is a direct consequence of architectural masking. By design, the platform’s interface facilitates a simplified view of reality for the general public, while burying the audit trail of capital influence within the complex, low-legibility layers of the blockchain. As we observed in Stage 3, the transparency of the system is asymmetric. While the blockchain is technically open for anyone to see, the literacy required to decipher "Whale" movements or identify "Wash Trading" creates a functional barrier to entry. This mirrors what sociotechnical scholars describe as Professionalization, where a supposedly public resource (in this case, "the truth" about the future) is captured by specialists who possess the tools to see through the "laundered" signals.

This capture fundamentally undermines the "Wisdom of the Crowd" theory. If the majority of users are not contributing independent beliefs but are instead reacting to the momentum of a few significant wallets, the market is no longer an aggregator of diverse perspectives. Instead, it becomes a feedback loop of capital. In this environment, the "consensus" is not discovered, but manufactured by those with the financial resources to move the price and the pseudonymity to hide their motives. By laundering this capital-driven influence into a clean, probabilistic percentage, the platform grants "Whale" interests the unearned authority of a democratic mandate.

Consequently, the resulting data is not a neutral reflection of public sentiment. It is a laundered signal that has been stripped of its financial origins. When policymakers or news organizations cite these percentages as "objective" forecasts, they are inadvertently participating in the final stage of the laundering process. They are granting institutional legitimacy to a signal that was forged in a state of extreme capital asymmetry, effectively allowing private wealth to masquerade as collective intelligence.

\subsection{The Accountability Gap: Laundering Moral and Institutional Responsibility}

A critical consequence of the laundering life cycle is the creation of an accountability gap. By the time a prediction reaches its final "hardened" state, the platform’s architecture has successfully decoupled the resulting data from the human and ethical costs associated with its production. This process mirrors the final stage of financial laundering integration, where the "cleaned" product is merged into the legitimate economy. In the context of prediction markets, this integration allows both the platform and its users to evade moral and institutional responsibility by attributing controversial outcomes to the "objective" mechanics of the market.

This gap is primarily sustained through "Moral Displacement", a phenomenon we observed in Stage 4. When ethically sensitive events, such as military conflicts or political instability, are re-framed as speculative puzzles, the interface performs a form of ethical sanitization. The "Laundered Prediction" allows the user to profit from tragedy without the social stigma of being a disaster speculator. Because the interface presents a clinical, stock-like chart rather than the human reality of the event, the trader can project the moral burden onto the market itself. As our participants noted, the platform "takes the responsibility of being a bad person" away from the individual, effectively commodifying human suffering into a neutral data point.

Further, the Oracle Fallacy provides the platform with institutional deniability. By offloading the "dirty labor" of truth-resolution to decentralized oracles like UMA, the platform can claim it has no hand in the final result. However, our audit shows that this "decentralization" is often a mirage. The high financial barriers—such as the \$750 bonds—ensure that only a specific class of actors can influence the final settlement. When a dispute arises, the platform can point to the "code" or the "oracle" as a neutral arbiter, even if the resolution was determined by capital-weighted voting. This creates a state of institutional evasion, in which the platform takes credit for the "accuracy" of its signals while distancing itself from the contested and often biased methods used to produce them.

This lack of accountability has profound implications for how these signals are consumed by the public. When news organizations or policy-makers treat a "hardened" Polymarket percentage as a factual forecast, they are interacting with a signal that has been scrubbed of its moral and tactical friction. The laundering is complete when the "dirty" history of Discord disputes, exit-liquidity tactics, and financial bullying is erased, leaving only a "clean" number that appears to be beyond human manipulation. This results in a dangerous epistemic loop: the market produces a signal through capital-driven bias, the platform hardens it into "truth", and society consumes it as a neutral fact, with no one held responsible for the underlying distortion.

\subsection{Algorithmic Gatekeeping and the Curated Future}
The laundering process begins with an act of ontological power which is the ability to define the boundaries of what is considered "bet-able" and, by extension, what is "knowable". While prediction markets are often theorized as open, permission-less mirrors of reality, our findings in Stage 1 reveal a rigorous system of algorithmic gatekeeping. By centralizing the creation of contracts within a "Markets Team", the platform does not merely aggregate existing information, but actively scripts the narrative of the future. This curation ensures that the only versions of reality allowed to reach the flattening stage are those that are measurable, binary, and compatible with blockchain settlement.

This gatekeeping functions as a form of epistemic enclosure. When a platform decides which questions are valid (e.g., "Who will win the election?") and which are "too vague" (e.g., "Will the community be better off?"), it effectively prioritizes data-centric reality over complex social lived experiences. As our participants observed, this creates a "curated set" of futures that fit a specific brand. This curation is the first step in the laundry: it scrubs away the messy, non-binary complexities of social life—such as "why" an event matters, and replaces them with a clean, tradable ontology. The platform essentially manufactures a reality that is pre-formatted for speculation, ensuring that the "wisdom" produced is always limited by the platform's own institutional interests.

This curation also creates a false neutrality. Because the front-end interface is clean and automated, users often perceive the available markets as a neutral reflection of global interest. However, our audit shows that the "Markets Team" acts as a primary editorial board, deciding which geopolitical tensions or social crises are worthy of being turned into financial instruments. This is not a passive reflection of the world, but an active commodification of uncertainty. By selecting specific events to be "hardened" into data, the platform exercises a subtle but immense influence over public attention and political discourse.

Ultimately, this gatekeeping ensures that the "laundering" is successful from the very start. By the time a trader enters the market, the complex "dirty" reality of a social event has already been converted into a "clean" binary contract. The platform has already decided the terms of the debate, leaving the user to simply provide the capital that will eventually be laundered into "truth". In our final discussion section, we explore how to break this cycle by introducing design interventions that restore the very friction that the laundering process seeks to erase.

\subsection{Design Recommendations: Toward Friction-Positive Interfaces}
The laundering process we have described relies on the "smoothing" of information by stripping away context, motives, and disputes to produce a clean, authoritative signal. To counter this, we propose a shift toward "Friction-Positive Design". "Positive Friction" refers to deliberate design choices that slow the user down, encouraging critical reflection rather than passive consumption \cite{chen2024exploring}. For prediction markets, this means re-inserting the very "noise" that the platform currently seeks to launder. We suggest three primary design interventions to restore epistemic integrity:

\subsubsection{\textbf{Exposing Capital Asymmetry}}
To break the "Architectural Masking" of Stage 3, platforms should implement "Whale Alerts" and "Concentration Metrics" directly on the front-end. If a 70\% probability is being driven by only three large wallets, the interface should visually flag this concentration. By moving this data from the specialist layer of block explorers to the public UI, the platform would dismantle the two-tiered hierarchy of Epistemic Stratification.

\subsubsection{\textbf{Visualizing Interpretive Context}}
To counter the "Probabilistic Flattening" of Stage 2, interfaces should allow for "Multi-Signal Disclosure." Instead of a single percentage, the platform could display a "Sentiment Map" that distinguishes between conviction bets, financial hedges, and tactical noise. Disclosing the intent behind the capital—even through optional user tagging—would prevent the "epistemic dilution" that occurs when a hedge is mistaken for a prediction.

\subsubsection{\textbf{Audit-Ready Resolution Trails}}
To address the "Accountability Gap" and the "Oracle Fallacy" of Stage 4, the history of a market’s resolution must remain visible. Currently, once a market settles, the "messy" history of Discord disputes and Oracle votes is scrubbed. A friction-positive design would archive and link this Dispute Log directly to the final result. This ensures that the "hardened" fact remains connected to its contested origin, preventing the platform from evading institutional responsibility for controversial settlements.

By embracing these "programmed inefficiencies", platforms can move from being machines of "laundering" to tools of genuine Algorithmic Accountability. These interventions do not seek to make the market faster or more efficient; rather, they seek to make it more honest. By exposing the friction behind the signal, we can ensure that prediction markets serve as a transparent mirror of collective intelligence rather than a sanitized megaphone for capital-driven influence.
\section{Limitations and Future Work}

This audit provides a deep qualitative mapping of the prediction laundering lifecycle, prioritizing interpretive depth over statistical generalizability. While our sample ($N=27$) included diverse experts and generalists, it does not represent the global user base or novices who consume these signals solely through news media. Furthermore, the rapid evolution of blockchain governance and interfaces may shift even as our structural logic remains applicable. Future work should combine sociotechnical analysis with on-chain data forensics and quantitative methods to better understand how hedge wagers and financial noise are technically converted into authoritative signals.

\section{Ethics and Positionality}
This study was conducted with the approval of the authors' Research Ethics Board (REB) to ensure the protection of participant anonymity and data security. Given the pseudonymous and often volatile nature of decentralized prediction markets, we took specific measures to ensure that participants were not encouraged to engage in real-world financial risk during the sessions. Furthermore, we recognize our role as researchers in a sociotechnical audit as one of non-participant observation. We maintained a deliberate outsider stance during the digital ethnography phase to avoid influencing market sentiment or community discourse on X and Discord. This positionality allowed us to document the platform's social and technical friction as it naturally occurs, providing a neutral baseline for our analysis of user interpretation.

\section{Generative AI Usage Statement}

The authors used Google Gemini and ChatGPT to assist in the preparation of this manuscript. Specifically, these generative AI models were used for refining the linguistic clarity and academic tone of the paper. All core concepts including the "Prediction Laundering" framework, the four stages of the laundering lifecycle, and the design recommendations were conceptualized by the human authors based on original empirical research. The authors reviewed, edited, and verified all outputs produced by the generative AI to ensure accuracy and adherence to academic standards. The authors take full responsibility for the content of this publication.

\bibliographystyle{ACM-Reference-Format}
\bibliography{sample-base}

\end{document}